\documentclass[twocolumn,prb,floatfix]{revtex4}
\usepackage{graphicx}
\usepackage{amsmath}
\usepackage{bm}

\begin{document}

\title[Roughening with elastic interaction]{Thermal roughening of an 
  SOS-model with elastic interaction}

\author{Frank Gutheim}
\author{Heiner M\"uller-Krumbhaar}
\author{Efim Brener}
\affiliation{Forschungszentrum J\"ulich, D-52425 J\"ulich, Germany}

\author{Vladimir Kaganer}
\affiliation{Paul-Drude-Institut f\"ur Festk\"orperelektronik, 
Hausvogteiplatz 5-7, D-10117 Berlin, Germany}

\begin{abstract}
  We analyze the effects of a long-ranged step-step interaction on 
  thermal roughening within the framework of a solid-on-solid model 
  of a crystal surface by means of Monte Carlo simulation. 
  A repulsive step-step interaction is modeled by elastic dipoles 
  located on sites adjacent to the steps. In order to reduce the 
  computational effort involved in calculating interaction energy
  based on long-ranged potentials, we employ a multi-grid 
  scheme. As a result of the long-range character of the step 
  interaction, the roughening temperature increases drastically 
  compared to a system with short-range cutoff as a consequence 
  of anti-correlations between surface defects.
\end{abstract}

\maketitle

\section{Introduction}

At low temperatures crystal surfaces are known to assume the shape of a 
plane facet. 
With increasing temperature fluctuations gradually 
contribute a nonzero thickness to the initially flat facet.
This surface thickness finally diverges at a finite temperature, 
the roughening temperature, 
where the order of the facet is lost completely.
This transition can be described by a set of 
renormalization group equations first analyzed by 
Kosterlitz and Thouless \cite{kosterlitz73:_order}. Because of its unusual 
properties and the relation to the two-dimensional 
Coulomb gas \cite{chui76:_phase_coulom}, 
this roughening transition has attracted substantial attention 
\cite{kosterlitz74,kosterlitz77:_coulom,%
ohta78:_renor_group_theor_inter_rough_trans,knops80:_momen_gordon}. 

Various discrete solid-on-solid models have been shown to undergo 
this type of transition.
Most of these models incorporate local interactions, at most
next-nearest neighbor interactions. Within some of these models 
a transition involving in-plane disorder is possible, 
usually referred to as preroughening 
\cite{prasad98:_layer,bastiaansen96:_rough,%
jagla99:_surfac_meltin_induc_preroug,%
noh97:_preroug_si_ge,%
woodraska97:_rough_preroug_diamon_cubic_surfac,%
prestipino95:_preroug_diffus_growt_surfac,%
nijs90:_preroug_cryst_surfac_energ_differ,%
mazzeo94:_rough}.  
Step-step interaction by means of elastic deformation of the crystal, however, 
is of a long-ranged nature and has apparently 
not been previously studied in the context of roughening.
Leaving the matter of preroughening aside, we will try to elucidate 
the effects of long-range elastic interactions on the roughening process.

The paper is organized as follows. First we will introduce elastic 
interaction between surface defects and suggest some simplifications 
to make the problem tractable. 
Then we present the details of our discrete 
solid-on-solid model allowing for long-range step interaction.
We will show the results of our extensive Monte Carlo simulations 
and interpret the effects.

\section{Step Interaction}

Elastic step interaction on the surface of a semi-infinite crystal 
can be described in terms of elastic force dipoles located at the step 
edges \cite{andreev81:_capil,marchenko80:_elast,hardy67,kaganer01:_energ}. 
Knowing the Green function $G_{ij}$ for an infinite elastic half-space 
one is able to calculate the elastic displacement field $u_{i}(\bm{r})$ 
from a given force density $f_{i}(\bm{r})$
\begin{equation}
    u_{i}(\bm{r})= \int d^{2}r'\, G_{ij}(\bm{r}-\bm{r'})\,f_{j}(\bm{r'}).
\end{equation}
The elastic energy $E_{\text{el}}$ becomes
\begin{eqnarray}
    E_{\text{el}} &=& -\int d^{2}r\, u_{i}(\bm{r})\,f_{i}(\bm{r}) \nonumber \\
    &=& - \int\int d^{2}r d^{2}r'\, G_{ij}(\bm{r}-\bm{r'})
    \,f_{j}(\bm{r'})\,f_{i}(\bm{r}).
\end{eqnarray}
Using that forces $f_{i}(\bm{r})$ are present only in the vicinity 
of a step and that the mono-pole 
moment at the step vanishes, we can rewrite the energy using 
force dipole densities $q_{ik}(\bm{r})$ as the next term in a 
multi-pole expansion
\begin{equation}\label{eq:integrand}
  E_{\text{el}} \approx 
    - \int\int d^{2}r d^{2}r'\,q_{jk}(\bm{r'})\,q_{il}(\bm{r}) 
    \partial_{k}\partial_{l}G_{ij}(\bm{r}-\bm{r'}) .
\end{equation} 

Using symmetry arguments one can determine two types of force dipoles 
that are considered to be present at a step \cite{marchenko80:_elast}. 
One type involves in-plane forces perpendicular to the step, 
the other arises from forces orthogonal to the crystal surface.
Due to the structure of the Green function, dipole tensors involving 
forces orthogonal to the surface show a behavior different from those 
involving only in-plane forces 
\cite{marchenko80:_elast,kaganer01:_energ}. 
The former lead to attractive or repulsive 
interaction depending on the signs of the steps, the latter 
produce a sign independent behavior, which is strictly repulsive. 
There are materials \cite{poon90:_equil_si,poon92:_ledge_si} 
where the sign dependent 
contributions are small compared to the step repulsion caused by 
in-plane forces, and we will restrict our model to the case, where we can 
neglect sign dependence of the steps.

In the case of isotropic linear elasticity the half-space elastic  
Green function $G_{ij}(\bm{r})$ can be written in a 
simple form \cite{andreev81:_capil}
\begin{equation}
 G_{ij}(\bm{r})= \frac{1+\sigma}{\pi E}\,  \frac{1}{r} 
 \left\{ 
   (1-\sigma)\delta_{ij} +\sigma \frac{r_{i}r_{j}}{r^{2}}
 \right\}
\end{equation}
where $i$ and $j$ are restricted to in-plane coordinates. 

For a step stretching in $y$-direction one would assume the force dipole 
tensor at the step to be of the type 
$q_{ij} \sim \delta_{ix}\delta_{jx}$. 
This means that the interaction between two line elements will depend on 
their orientation. 

 In the case of two parallel steps, a distance $d$ in $y$-direction 
apart, the interaction energy density $w$ (per area squared) 
can be computed by evaluating the integrand from 
eq.~(\ref{eq:integrand}) for two interacting force dipoles of the type
$q_{ij}= \delta_{ix}\delta_{jx}$. 
It is given by
\begin{eqnarray}
  w(r,\varphi)& =& \gamma\, \left[ \frac{3 \cos \varphi -1 }{r^{3}} 
      \right. \nonumber\\
         & +&  \left. \frac{\sigma}{1-\sigma} 
          \frac{2 + 15 \cos^{4}\varphi - 15\cos^{2}\varphi }{r^{3}}
        \right]\, \label{eq:angulardipolesamedirection}
\end{eqnarray}
where $\varphi$ denotes the angle between the radius vector $\bm{r}$ and the 
orientation of the dipole forces, which is given by 
$\varphi=\arctan(\Delta y / d)$, and $\Delta y$ is the distance between 
the dipoles in $y$-direction.
The factor $\gamma$ is given by
\begin{equation}
  \gamma=\frac{1-\sigma^{2}}{\pi E}\, \tilde Q^{2},
\end{equation}
where $\tilde Q=Q/a$ is the dipole moment (per unit step length) 
and $Q$ would be the dipole moment assigned to a single atom at the step 
edge. 

Integrating the energy density for a configuration with two parallel steps at 
distance $d$, we state that the energy per unit length of the line is just
\begin{equation}
  \tilde W =  4\,\gamma\, \frac{1}{d^{2}}
   -2\,\gamma\, \frac{1}{\varepsilon^{2}}\,\frac {1-2\sigma}{1-\sigma},
\end{equation}
where the interaction was limited to distances greater $\varepsilon$.
Note that the second term, which contributes to line energy, is 
negative for all possible Poisson ratios $-1\le\sigma\le 1/2$.

In order to make another simplification of the step-step interaction we 
compare the above result to the case of a scalar $w\sim 1/r^{3}$ interaction 
associated with isotropic dipoles $q_{ij}\sim \delta_{ij}$,
\begin{equation}\label{eq:scalarint}
  \tilde W_{\text{scalar}} =  4\,\gamma\, \frac{1}{d^{2}}
   +2\,\gamma\, \frac{1}{\varepsilon^{2}},
\end{equation}
from which we conclude that the only difference in this specific geometry
is a change in the line energy, which is mainly due to contributions from 
short range interactions.

Because we aim at showing the effect of long-range interactions 
on the thermal roughening process, we neglect the angular dependence 
completely 
and assume that the dipole moments are isotropic. 
This leads to a simple isotropic $1/r^{3}$-interaction 
between force dipoles. 
Furthermore this ensures that the elastic 
contribution to the step energy is positive.

\section{Model Description}

\begin{figure}[tbh]
\begin{center}
\includegraphics[width=0.85\linewidth]{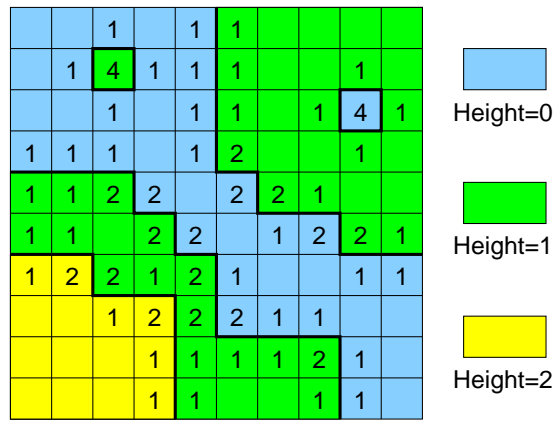}
\caption{The numbers of dipole charges assigned to a lattice sites is 
  proportional to the accumulated absolute height difference
  corresponding to eq.~\ref{eq:totalheightdifference}.
  \label{fig:qcharges}}
\end{center}
\end{figure}
Within the framework of a solid-on-solid model we describe the crystal 
surface by a simple height field of integer numbers $h$. 
Like in a common SOS model, overhangs are forbidden.
Instead of the usual surface energy term (summation over nearest neighbors)
\begin{equation}
  E_{\text{surf}} = \frac{J}{2}\sum\limits_{<i,j>} |h_{i}-h_{j}|^{\alpha},
\end{equation}
with coupling constant $J$ and $\alpha=1$, $2$ for the ASOS-model and the 
DGSOS-model respectively, 
we define an elastic step interaction by introducing a field of 
elastic dipole charges $q$. To every lattice site a dipole charge $q_{k}$
proportional to the number of height differences to the four  
neighboring sites is assigned, i.\,e.\ site $k$ carries a number of
\begin{equation}\label{eq:totalheightdifference}
q_{k}=\frac{1}{2}\sum\limits_{<i,j>} |h_{i}-h_{j}| \delta_{ik}
\end{equation}
charges. Figure~\ref{fig:qcharges} gives an example how charges are assigned 
to a simple height field configuration. The elastic dipole charges 
interact, in consequence of eq.\ (\ref{eq:scalarint}), 
via a modified $r^{-3}$ interaction potential $\Psi(r)$,
\begin{equation}\label{eq:psidef}
  \Psi(r)=  \left\{ 
      \begin{array}{cl}
        r^{-3}  & \quad \text{if } r \ge 1 \\
        1       & \quad \text{if } r = 0
      \end{array}  
    \right. ,
\end{equation}
where $r$ is the in-plane distance between two lattice sites 
measured in units of the lattice constant. 
This gives rise to the elastic energy 
\begin{equation}\label{eq:elasticsum}
  E_{\text{el}} = \frac{w_{\text{el}}}{2}\sum\limits_{i,j} 
   q_{i}q_{j}\Psi(r_{ij}),
\end{equation} 
where $r_{ij}$ is the distance between lattice sites $i$ and $j$ and 
$w_{\text{el}}$ 
can be adjusted to give the desired interaction strength. Note that the 
case $i=j$ is not excluded from the summation.
During our simulation this constant had the value $w_{\text{el}}=0.31$, which 
gives a line energy of about $0.5$ per unit length for infinite range 
interaction. 

Later we also limit the range of interaction. For this purpose 
we introduce a cutoff-potential $\Psi_{l}$ with cutoff length $l$
\begin{equation}
  \Psi_{l}(r)= \left\{ 
      \begin{array}{cl}
        \Psi(r)   & \quad \text{if } r \le l \\
        0      & \quad \text{if } r>l
      \end{array}  
    \right.
\end{equation}
which vanishes for distances greater than $l$.

For two straight steps of length $L$ with distance $d$ this 
elastic energy contribution consists of the self energies of the steps 
and the expected $\sim d^{-2}$ step interaction term
\begin{equation}
  E_{\text{int}}\sim \frac{L}{d^{2}}.
\end{equation}
The self energy 
contribution of a straight step can be adjusted to be the same as the
line energy of a DGSOS model. 
Because in this model both the line 
energy and the step interaction originate from eq.~(\ref{eq:elasticsum}), 
their relative amplitude is fixed and we can concentrate on the crossover 
from a local to a long-range model depending on the cutoff length $l$.
Note, however, that other relative amplitudes can in principle be obtained
by a different choice of $\Psi(r)$ at small distances $r<1$ in 
eq.~(\ref{eq:psidef}).

The simulation is carried out on a square lattice of size 
$L^{2}=64\times 64$ to $128\times 128$. 
In order to calculate the difference in energy for 
every metropolis Monte Carlo trial, we apply a multi-grid 
scheme based on \cite{steinbrecher99:_fract} which has already been applied 
successfully to submonolayer epitaxy \cite{gutheim01:_epitax}.

This cuts down the computational 
costs from order $L^{4}$ to order $L^{2}\log(L)$ for each time-step, 
which has to be multiplied by an additional factor of $L^{2}$, for the number
of time-steps the system needs to equilibrate. Without the use of the 
multi-grid scheme the computational costs would not have permitted system sizes
beyond $L=25$. Still the system size, $L\le 128$, 
is rather restricted and we are aware that the results should be accounted
as qualitative rather than quantitative. However, computations on the DGSOS 
and ASOS models at $L=128$, which we did for comparison, 
give transition temperatures $T_{\text{R}}\approx 1.5 J$ and 
$T_{\text{R}}\approx 1.25 J$ 
respectively, which agree reasonably well with known 
results \cite{lapujoulade94}.

\section{Results}
\subsection{Height Correlation Function}
\begin{figure}%[tbh]
\begin{center}
\includegraphics[width=0.85\linewidth, clip=true]{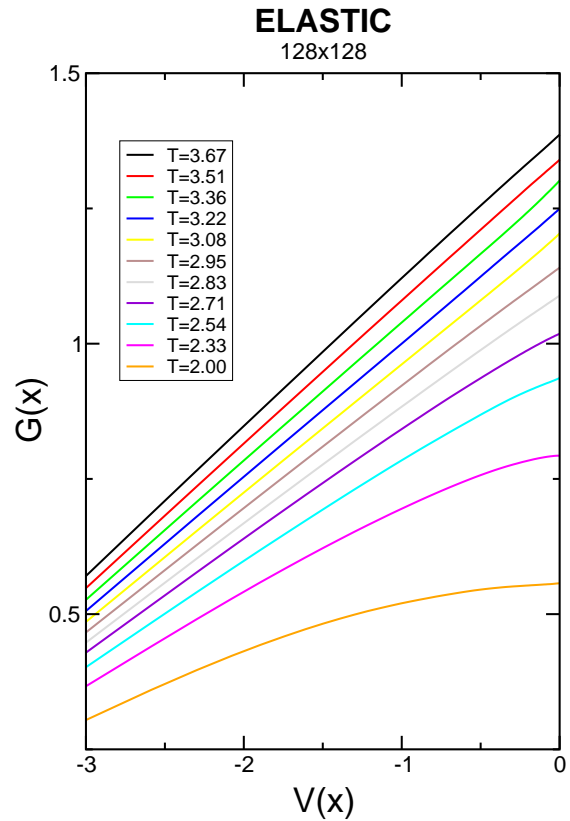}
\caption{Height-height correlation function without cutoff. 
The correlation function saturates for small temperatures and 
shows logarithmic behavior for $T>T_{\text{R}}$. The first straight 
line gives an estimate of $T_{\text{R}}\approx 2.8$. 
\label{fig:correlast}} 
\end{center}
\end{figure}

\begin{figure}%[tbh]
\begin{center}
\includegraphics[width=0.85\linewidth, clip=true]{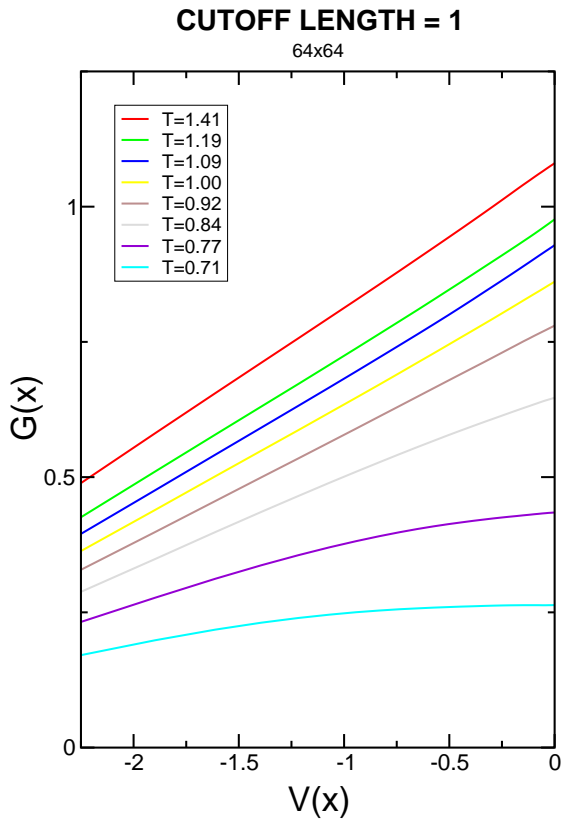}
\caption{Height-height correlation function with a cutoff length 1, i.e. only 
charges on nearest-neighbor-sites interact. The first straight 
line gives an estimate of $T_{\text{R}}\approx 0.9$.\label{fig:length1}} 
\end{center}
\end{figure}

We determine the roughening temperature $T_{\text{R}}$ 
from the behavior of the height-height correlation function. 
Below roughening, $T<T_{\text{R}}$, the 
interface is macroscopically flat, i.\,e.\ 
the height-height correlation function 
\begin{equation}
  G(r) =  \langle [h(0)-h(r)]^{2} \rangle
\end{equation}
approaches a finite value in the limit $r\to \infty$. To be more precise, the 
correlation length $\xi$ is finite and the interface has a characteristic 
width. Approaching the transition temperature
the correlation length increases and diverges at $T=T_{\text{R}}$. 
For $T>T_{\text{R}}$ the correlation function $G(r)$ diverges 
\cite{lapujoulade94} according to the conventional theory of the roughening 
transition,
\begin{equation}
  G(r) \sim K(T)\log r,
\end{equation}
with an amplitude $K(T)$ depending on the temperature.
Plotting $G(r)$ vs.\ $\log r$ one could determine at what temperature 
the correlation length $\xi$ diverges and the graphs approach a straight 
line. 
 
In a finite system with periodic boundary conditions, however, the 
correlation length $\xi$ cannot exceed the system size $L$, the height-height
correlation function $G(r)$ saturates for $T>T_{\text{R}}$ as well. 
In order to overcome this finite-size effect, we will use an approach 
similar to the one used in \cite{saito81:_two_coulom}. In order to keep the 
argument simple we only consider correlations along the main directions 
of the lattice and replace $r$ by $x$. 

As the limiting behavior of $G(x)$ for periodic boundary conditions 
has to be a periodic function that behaves like the logarithm for 
distances $\ll L$, we define a ``periodic logarithm'' by means of Fourier 
analysis. 
In order to avoid the singularity at $r\to 0$ we start with 
\begin{equation}
  v(x) = \max( \log(x), 1)
\end{equation}
and the integral-Fourier or, using symmetry arguments, the cosine transform
\begin{equation}
  \tilde v(k) = \frac{1}{\pi}\int\limits_{0}^{\infty} \cos(kx) v(x)dx.
\end{equation}
Making use of these Fourier components we define the $L$-periodic function 
$V_{L}(x)$
\begin{equation}
  V_{L}(x) = \frac{2\pi}{L} \sum\limits_{n=1}^{\infty} 
  \tilde v\left(\frac{2\pi n }{L}\right) 
  \cos\left( \frac{2\pi n }{L} x \right) 
    \frac{\sin\left(\frac{2\pi n }{L}\right)}{\frac{2\pi n }{L}} ,
\end{equation}
which is a discrete back transform averaged over unit distances. For 
convenience we define
\begin{equation}
  V(x) = V_{L}(x)-V_{L}(L/2)
\end{equation}
and plot $G(x)$ vs.\ $V(x)$. 
Fig.~\ref{fig:correlast} shows the correlation function for the case 
of the full $1/r^{3}$ interaction. At a temperature of about $T=2.8$ 
the graph becomes straight, indicating the roughening transition.
Restricting the elastic dipole charge interaction to distances $\le 1$, 
the graph of the correlation function becomes straight at a lower 
temperature $T=0.9$, see Fig.~\ref{fig:length1}.

\begin{figure}%[tbh]
\begin{center}
\includegraphics[width=0.80\linewidth,clip=true]{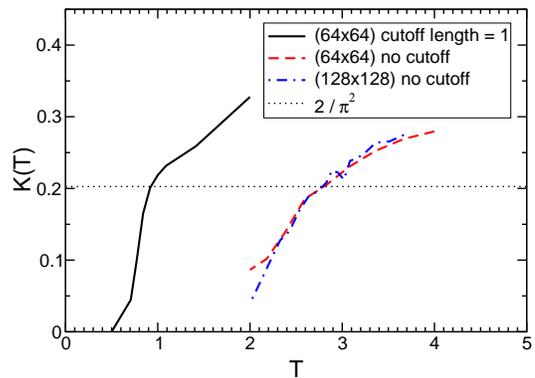}
\caption{Slope $K(T)$ vs.\ temperature. Estimation of $T_{\text{R}}$ using 
the universal value from conventional roughening theory gives $T_{\text{R}}=2.8$ 
for infinitely ranged interaction and $T_{\text{R}}=0.9$
for cutoff length 1.\label{fig:slope}}
\end{center}
\end{figure}
\begin{figure}%[tbh]
\begin{center}
\includegraphics[width=0.80\linewidth,clip=true]{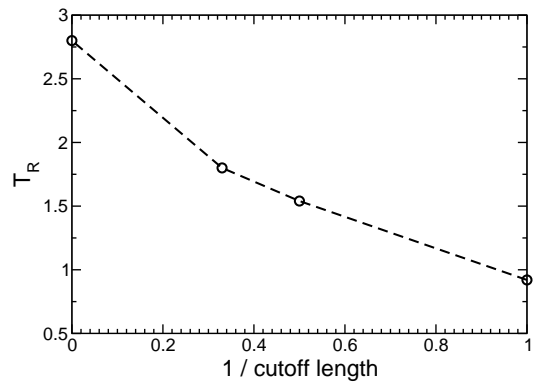}
\caption{Roughening temperature $T_{\text{R}}$ vs.\ 
inverse cutoff length $1/l$.
Even at $l=3$ the roughening temperature 
$T_{\text{R}}\approx 1.8$ is well below the infinite range potential 
value. \label{fig:trvscut}}
\end{center}
\end{figure}

From the Kosterlitz-Thouless theory of the roughening transition, the slope 
of the correlation function 
is expected to assume the universal 
value $K(T_{\text{R}})=2/\pi^{2}$. 
Plotting slope vs.\ temperature one can also obtain an estimate of the 
roughening temperature, see Fig.~\ref{fig:slope}. 
From this we obtain identical estimates for the two cases with or without 
cutoff.

From this we conclude that the system with long-range interaction has a 
much higher transition temperature compared to the model with interaction 
cutoff, the roughening temperature changes by a factor $\sim 3$.

Note that the roughening temperature increases gradually with the cutoff 
length, see Fig.~\ref{fig:trvscut}. Even at $l=3$ the roughening temperature 
$T_{\text{R}}\approx 1.8$ 
is still well below the value for infinite range interaction. The 
increase of the roughening temperature is not a next nearest neighbor 
effect.

\subsection{Energetic Scales}

One might argue that increasing the range of the interaction potential 
just changes the relevant energetic scale. However, the energetic scales 
one is usually tempted to think of, i.e. the energy of a straight step or single 
kinks on such a step, do not change by more than $36\%$. The straight line 
energy per unit step length increases from $w_{\text{l}}=0.39$ to 
$0.53$, the corresponding kink energy changes 
from $w_{\text{k}}=0.88$ to $1.18$. 
In the low temperature regime, the energy of one single ad-atom on a flat 
crystal surface is the important energetic scale, which changes from 
$w_{\text{a}}=2.53$ to $2.69$, an increase by no more than $6\%$. 

It should be noted that the main contribution to the change of these
energetic scales comes form short-range interactions. Using a cutoff length of 
$l=3$, the straight line and kink energies are only about $6-7\%$ below the 
the full potential value, whereas the single ad-atom defect 
energy deviates by no more than $0.05\%$.

From the change of these energetic scales one usually would expect an 
equal increase of the roughening temperature. One would hesitate, however, 
to make these changes responsible for an increase of the roughening 
temperature by a factor of $\sim 3$.

\subsection{Average Energy}

\begin{figure}%[tbh]
\begin{center}
\includegraphics[width=0.85\linewidth,clip=true]{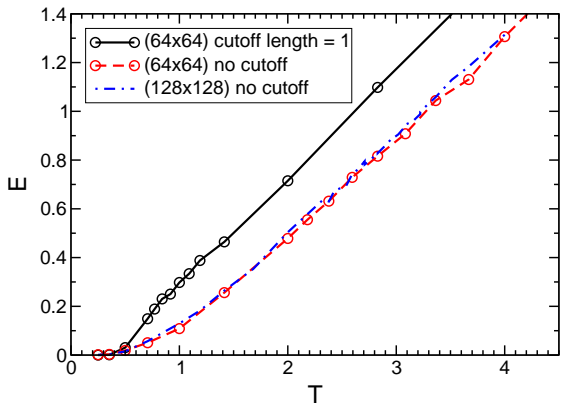}
\caption{Average energy per unit surface $E$ vs.\ temperature $T$. 
  Average energy for the cutoff-potential is strictly higher in comparison
  to the long-range case.\label{fig:avenerg}}
\end{center}
\end{figure}

\begin{figure}%[tbh]
\begin{center}
\includegraphics[width=0.85\linewidth,clip=true]{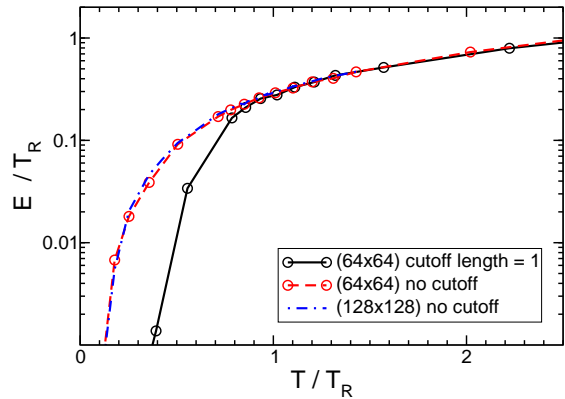}
\caption{Scaled average energy per unit surface $E$ vs.\ scaled 
temperature $T/T_{\text{R}}$. For $T/T_{\text{R}}\ge 1$ the scaled 
data collapses onto a single graph.\label{fig:avenergscaled}}
\end{center}
\end{figure}

\begin{figure}%[tbh]
\begin{center}
\includegraphics[width=0.85\linewidth,clip=true]{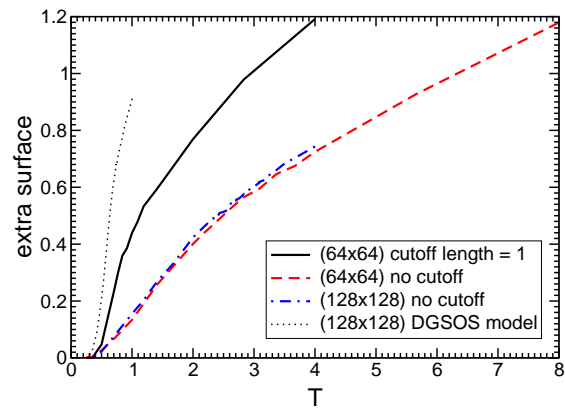}
\caption{Extra surface, i.\,e.\ number of broken bonds, 
  vs.\ temperature $T$. Fewer defects are created when no cutoff is 
  used. \label{fig:surface}}
\end{center}
\end{figure}

Comparing the average energy $E$ of the system 
computed with and without restriction of the charge interaction 
range, one clearly sees that the energy for the non restricted 
interaction always stays well below the graph of the restricted 
system, see Fig.~\ref{fig:avenerg}. For high temperatures the average energy 
$E$ goes linear with temperature $T$, indicating that the heat capacity 
becomes constant. 

The range of the interaction potential only affects the behavior below 
the transition temperature. Above the roughening transition all details 
of the interaction are combined into one single parameter, 
the roughening temperature $T_{\text{R}}$. Accordingly the scaled graphs 
$E/T_{\text{R}}$ vs.\ $T/T_{\text{R}}$  
coincide for $T/T_{\text{R}}>1$, see Fig.~\ref{fig:avenergscaled}.

The decrease in average energy of the system using long-ranged interaction
coincides with a smaller number of broken bonds, see 
Fig.~\ref{fig:surface}. The number of deviations from a facet or the 
step length is smaller compared to the system with interaction potential 
cutoff. 

\subsection{Defect Correlations}

\begin{figure}
\begin{center}
\includegraphics[width=0.85\linewidth,clip=true]{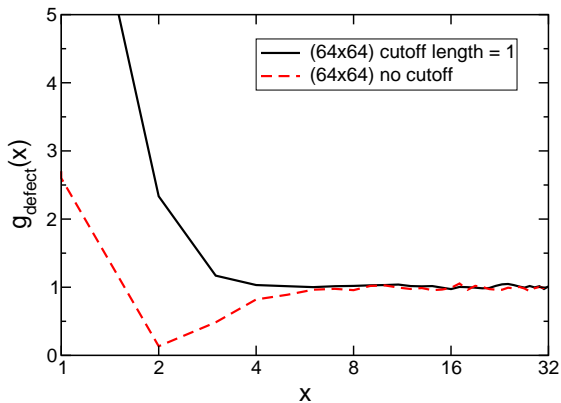}
\caption{Defect correlation at $T=0.5$ for both long-range interaction 
and cutoff. The long-range interaction causes a stronger repulsion gap 
(here near $x=2$), which means that the defects prefer to be separated. 
This anti-correlation effect is 
responsible for a strong decrease in entropy.
\label{fig:defectcorr}}
\end{center}
\end{figure}

Restricting the surface height to $\{-1,0,+1\}$, one may talk about a 
defect wherever the height deviates from the average height $0$. Then one 
can analyze correlation between these defects, i.e. the thermal average of
\begin{equation}
  g_{\text{defect}}(r) = \frac {\langle 
    [ h^{2}(\bm{r'})-h^{2}(\bm{r'}+\bm{r}) ]^{2}\rangle_{\bm{r'}} }{
  \langle h^{2}(\bm{r'})\rangle_{\bm{r'}} ^{2}}.
\end{equation}
Scaled like this, the defect correlation 
will approach the value $1$ for large distances $r$. 
At low temperatures repulsion between the defects causes the graph to 
fall below value $1$ at midrange distances and ends well above value $1$ 
at distance $r=1$, because contact between equal defects is favored due 
to what might be called surface or line energy. Increasing the temperature, 
this repulsion gap will become smaller and vanish eventually. 

Fig.~\ref{fig:defectcorr} 
shows the defect correlation for both long-range interaction 
and cutoff for identical temperature. Whereas for long-range interaction 
the gap is still present, it has already vanished from the system 
with cutoff. The pronounced repulsion gap in the case of the infinite 
range interaction means that a single defect or island avoids being close 
to other defects. This cuts down the number of favorable configurations 
and thus reduces the entropy contribution to the free energy for 
given density of defects $n$. 

For the following argument we will assume 
that the main result is a reduction of entropy by some factor $\alpha<1$, 
whereas the average energy at given $n$ remains unchanged.
In a rather simplified picture we can then write the free energy 
like $F_{\alpha}=E(n)-T \alpha S(n)$, 
where $n$ depends on temperature $T$ and is determined 
by $\partial F / \partial n = 0$.
In this picture the free energy $F_{\alpha}(T)$ 
of the system with reduced entropy at temperature $T$ 
has the same properties as the original system at a 
lower temperature $\alpha T$. Thus if the original system 
had a roughening temperature $T_{R}$ the transition temperature 
$\tilde T_{R}$ of the system with reduced entropy will increase to 
$\tilde T_{R}=T_{R}/\alpha$.

\section{Conclusion}

In summary, we have presented a model which contains the essential effects 
of long-range elastic repulsion between steps on a crystal surface.
We conclude that correlations due to these 
long-range interactions can strongly 
increase the roughening temperature in solid-on-solid models, mainly by 
a reduction of the entropy. Since defects prefer to exist 
in secluded areas, the number of favorable configurations and consequently 
the entropy contribution to the free energy is diminished, leading to 
an increase of the roughening temperature. 
Our simulations suggest that the type of transition remains the same, 
although a 
rigorous proof lies beyond the scope of this type of Monte Carlo approach.

\begin{acknowledgments}
The authors would like to thank V.\,I.\ Marchenko for valuable 
comments and suggestions.
\end{acknowledgments}

\bibliography{preamble,submono,roughening}

\begin{thebibliography}{24}
\expandafter\ifx\csname natexlab\endcsname\relax\def\natexlab#1{#1}\fi
\expandafter\ifx\csname bibnamefont\endcsname\relax
  \def\bibnamefont#1{#1}\fi
\expandafter\ifx\csname bibfnamefont\endcsname\relax
  \def\bibfnamefont#1{#1}\fi
\expandafter\ifx\csname citenamefont\endcsname\relax
  \def\citenamefont#1{#1}\fi
\expandafter\ifx\csname url\endcsname\relax
  \def\url#1{\texttt{#1}}\fi
\expandafter\ifx\csname urlprefix\endcsname\relax\def\urlprefix{URL }\fi
\providecommand{\bibinfo}[2]{#2}
\providecommand{\eprint}[2][]{\url{#2}}

\bibitem[{\citenamefont{Kosterlitz and Thouless}(1973)}]{kosterlitz73:_order}
\bibinfo{author}{\bibfnamefont{J.~M.} \bibnamefont{Kosterlitz}}
  \bibnamefont{and} \bibinfo{author}{\bibfnamefont{D.~J.}
  \bibnamefont{Thouless}}, \bibinfo{journal}{J. Phys. C}
  \textbf{\bibinfo{volume}{6}}, \bibinfo{pages}{1181} (\bibinfo{year}{1973}).

\bibitem[{\citenamefont{Chui and Weeks}(1976)}]{chui76:_phase_coulom}
\bibinfo{author}{\bibfnamefont{S.~T.} \bibnamefont{Chui}} \bibnamefont{and}
  \bibinfo{author}{\bibfnamefont{J.~D.} \bibnamefont{Weeks}},
  \bibinfo{journal}{Phys. Rev. B} \textbf{\bibinfo{volume}{14}},
  \bibinfo{pages}{4978} (\bibinfo{year}{1976}).

\bibitem[{\citenamefont{Kosterlitz}(1974)}]{kosterlitz74}
\bibinfo{author}{\bibfnamefont{J.~M.} \bibnamefont{Kosterlitz}},
  \bibinfo{journal}{J. Phys. C} \textbf{\bibinfo{volume}{7}},
  \bibinfo{pages}{1046} (\bibinfo{year}{1974}).

\bibitem[{\citenamefont{Kosterlitz}(1977)}]{kosterlitz77:_coulom}
\bibinfo{author}{\bibfnamefont{J.~M.} \bibnamefont{Kosterlitz}},
  \bibinfo{journal}{J. Phys. C} \textbf{\bibinfo{volume}{10}},
  \bibinfo{pages}{3753} (\bibinfo{year}{1977}).

\bibitem[{\citenamefont{Ohta and
  Kawasaki}(1978)}]{ohta78:_renor_group_theor_inter_rough_trans}
\bibinfo{author}{\bibfnamefont{T.}~\bibnamefont{Ohta}} \bibnamefont{and}
  \bibinfo{author}{\bibfnamefont{K.}~\bibnamefont{Kawasaki}},
  \bibinfo{journal}{Prog. Theor. Phys.} \textbf{\bibinfo{volume}{60}},
  \bibinfo{pages}{365} (\bibinfo{year}{1978}).

\bibitem[{\citenamefont{Knops and Ouden}(1980)}]{knops80:_momen_gordon}
\bibinfo{author}{\bibfnamefont{H.~J.~F.} \bibnamefont{Knops}} \bibnamefont{and}
  \bibinfo{author}{\bibfnamefont{L.~W.~J.} \bibnamefont{Ouden}},
  \bibinfo{journal}{Physica A} \textbf{\bibinfo{volume}{103}},
  \bibinfo{pages}{597} (\bibinfo{year}{1980}).

\bibitem[{\citenamefont{Bastiaansen and Knops}(1996)}]{bastiaansen96:_rough}
\bibinfo{author}{\bibfnamefont{P.~J.~M.} \bibnamefont{Bastiaansen}}
  \bibnamefont{and} \bibinfo{author}{\bibfnamefont{H.~J.~F.}
  \bibnamefont{Knops}}, \bibinfo{journal}{Phys. Rev. B}
  \textbf{\bibinfo{volume}{53}}, \bibinfo{pages}{126} (\bibinfo{year}{1996}).

\bibitem[{\citenamefont{den
  Nijs}(1990)}]{nijs90:_preroug_cryst_surfac_energ_differ}
\bibinfo{author}{\bibfnamefont{M.}~\bibnamefont{den Nijs}},
  \bibinfo{journal}{Phys. Rev. Lett.} \textbf{\bibinfo{volume}{64}},
  \bibinfo{pages}{435} (\bibinfo{year}{1990}).

\bibitem[{\citenamefont{Jagla et~al.}(1999)\citenamefont{Jagla, Prestipino, and
  Tosatti}}]{jagla99:_surfac_meltin_induc_preroug}
\bibinfo{author}{\bibfnamefont{E.}~\bibnamefont{Jagla}},
  \bibinfo{author}{\bibfnamefont{S.}~\bibnamefont{Prestipino}},
  \bibnamefont{and} \bibinfo{author}{\bibfnamefont{E.}~\bibnamefont{Tosatti}},
  \bibinfo{journal}{Phys. Rev. Lett.} \textbf{\bibinfo{volume}{83}},
  \bibinfo{pages}{2753} (\bibinfo{year}{1999}).

\bibitem[{\citenamefont{Mazzeo et~al.}(1994)\citenamefont{Mazzeo, Jug, Levi,
  and Tosatti}}]{mazzeo94:_rough}
\bibinfo{author}{\bibfnamefont{G.}~\bibnamefont{Mazzeo}},
  \bibinfo{author}{\bibfnamefont{G.}~\bibnamefont{Jug}},
  \bibinfo{author}{\bibfnamefont{A.~C.} \bibnamefont{Levi}}, \bibnamefont{and}
  \bibinfo{author}{\bibfnamefont{E.}~\bibnamefont{Tosatti}},
  \bibinfo{journal}{Phys. Rev. B} \textbf{\bibinfo{volume}{49}},
  \bibinfo{pages}{7625} (\bibinfo{year}{1994}).

\bibitem[{\citenamefont{Noh and den Nijs}(1997)}]{noh97:_preroug_si_ge}
\bibinfo{author}{\bibfnamefont{J.~D.} \bibnamefont{Noh}} \bibnamefont{and}
  \bibinfo{author}{\bibfnamefont{M.}~\bibnamefont{den Nijs}},
  \bibinfo{journal}{J. Phys. A} \textbf{\bibinfo{volume}{30}},
  \bibinfo{pages}{7375} (\bibinfo{year}{1997}).

\bibitem[{\citenamefont{Prestipino et~al.}(1995)\citenamefont{Prestipino,
  Santoro, and Erio}}]{prestipino95:_preroug_diffus_growt_surfac}
\bibinfo{author}{\bibfnamefont{S.}~\bibnamefont{Prestipino}},
  \bibinfo{author}{\bibfnamefont{G.}~\bibnamefont{Santoro}}, \bibnamefont{and}
  \bibinfo{author}{\bibfnamefont{T.}~\bibnamefont{Erio}},
  \bibinfo{journal}{Phys. Rev. Lett.} \textbf{\bibinfo{volume}{75}},
  \bibinfo{pages}{4468} (\bibinfo{year}{1995}).

\bibitem[{\citenamefont{Woodraska and
  Jaszczak}(1997)}]{woodraska97:_rough_preroug_diamon_cubic_surfac}
\bibinfo{author}{\bibfnamefont{D.~L.} \bibnamefont{Woodraska}}
  \bibnamefont{and} \bibinfo{author}{\bibfnamefont{J.~A.}
  \bibnamefont{Jaszczak}}, \bibinfo{journal}{Phys. Rev. Lett.}
  \textbf{\bibinfo{volume}{78}}, \bibinfo{pages}{258} (\bibinfo{year}{1997}).

\bibitem[{\citenamefont{Prasad and Weichmann}(1998)}]{prasad98:_layer}
\bibinfo{author}{\bibfnamefont{A.}~\bibnamefont{Prasad}} \bibnamefont{and}
  \bibinfo{author}{\bibfnamefont{P.~B.} \bibnamefont{Weichmann}},
  \bibinfo{journal}{Phys. Rev. B} \textbf{\bibinfo{volume}{57}},
  \bibinfo{pages}{4900} (\bibinfo{year}{1998}).

\bibitem[{\citenamefont{Andreev and Kosevich}(1981)}]{andreev81:_capil}
\bibinfo{author}{\bibfnamefont{A.}~\bibnamefont{Andreev}} \bibnamefont{and}
  \bibinfo{author}{\bibfnamefont{Y.}~\bibnamefont{Kosevich}},
  \bibinfo{journal}{Sov. Phys. JETP} \textbf{\bibinfo{volume}{54}},
  \bibinfo{pages}{761} (\bibinfo{year}{1981}).

\bibitem[{\citenamefont{Marchenko and Parshin}(1980)}]{marchenko80:_elast}
\bibinfo{author}{\bibfnamefont{V.}~\bibnamefont{Marchenko}} \bibnamefont{and}
  \bibinfo{author}{\bibfnamefont{A.}~\bibnamefont{Parshin}},
  \bibinfo{journal}{Sov. Phys. JETP} \textbf{\bibinfo{volume}{52}},
  \bibinfo{pages}{129} (\bibinfo{year}{1980}).

\bibitem[{\citenamefont{Hardy and Bullough}(1967)}]{hardy67}
\bibinfo{author}{\bibfnamefont{J.}~\bibnamefont{Hardy}} \bibnamefont{and}
  \bibinfo{author}{\bibfnamefont{R.}~\bibnamefont{Bullough}},
  \bibinfo{journal}{Phil. Mag.} \textbf{\bibinfo{volume}{15}},
  \bibinfo{pages}{237} (\bibinfo{year}{1967}).

\bibitem[{\citenamefont{Kaganer and Ploog}(2001)}]{kaganer01:_energ}
\bibinfo{author}{\bibfnamefont{V.~M.} \bibnamefont{Kaganer}} \bibnamefont{and}
  \bibinfo{author}{\bibfnamefont{K.~H.} \bibnamefont{Ploog}},
  \bibinfo{journal}{Phys. Rev. B} \textbf{\bibinfo{volume}{64}}
  (\bibinfo{year}{2001}).

\bibitem[{\citenamefont{Poon et~al.}(1990)\citenamefont{Poon, Yip, Ho, and
  Abraham}}]{poon90:_equil_si}
\bibinfo{author}{\bibfnamefont{T.~W.} \bibnamefont{Poon}},
  \bibinfo{author}{\bibfnamefont{S.}~\bibnamefont{Yip}},
  \bibinfo{author}{\bibfnamefont{P.~S.} \bibnamefont{Ho}}, \bibnamefont{and}
  \bibinfo{author}{\bibfnamefont{F.~F.} \bibnamefont{Abraham}},
  \bibinfo{journal}{Phys. Rev. Lett.} \textbf{\bibinfo{volume}{65}},
  \bibinfo{pages}{2161} (\bibinfo{year}{1990}).

\bibitem[{\citenamefont{Poon et~al.}(1992)\citenamefont{Poon, Yip, Ho, and
  Abraham}}]{poon92:_ledge_si}
\bibinfo{author}{\bibfnamefont{T.~W.} \bibnamefont{Poon}},
  \bibinfo{author}{\bibfnamefont{S.}~\bibnamefont{Yip}},
  \bibinfo{author}{\bibfnamefont{P.~S.} \bibnamefont{Ho}}, \bibnamefont{and}
  \bibinfo{author}{\bibfnamefont{F.~F.} \bibnamefont{Abraham}},
  \bibinfo{journal}{Phys. Rev. B} \textbf{\bibinfo{volume}{45}},
  \bibinfo{pages}{3521} (\bibinfo{year}{1992}).

\bibitem[{\citenamefont{Steinbrecher et~al.}(1999)\citenamefont{Steinbrecher,
  M{\"u}ller-Krumbhaar, Brener, Misbah, and Peyla}}]{steinbrecher99:_fract}
\bibinfo{author}{\bibfnamefont{J.}~\bibnamefont{Steinbrecher}},
  \bibinfo{author}{\bibfnamefont{H.}~\bibnamefont{M{\"u}ller-Krumbhaar}},
  \bibinfo{author}{\bibfnamefont{E.}~\bibnamefont{Brener}},
  \bibinfo{author}{\bibfnamefont{C.}~\bibnamefont{Misbah}}, \bibnamefont{and}
  \bibinfo{author}{\bibfnamefont{P.}~\bibnamefont{Peyla}},
  \bibinfo{journal}{Phys. Rev. E} \textbf{\bibinfo{volume}{59}},
  \bibinfo{pages}{5600} (\bibinfo{year}{1999}).

\bibitem[{\citenamefont{Gutheim et~al.}(2001)\citenamefont{Gutheim,
  M{\"u}ller-Krumbhaar, and Brener}}]{gutheim01:_epitax}
\bibinfo{author}{\bibfnamefont{F.}~\bibnamefont{Gutheim}},
  \bibinfo{author}{\bibfnamefont{H.}~\bibnamefont{M{\"u}ller-Krumbhaar}},
  \bibnamefont{and} \bibinfo{author}{\bibfnamefont{E.}~\bibnamefont{Brener}},
  \bibinfo{journal}{Phys. Rev. E} \textbf{\bibinfo{volume}{63}}
  (\bibinfo{year}{2001}).

\bibitem[{\citenamefont{Lapujoulade}(1994)}]{lapujoulade94}
\bibinfo{author}{\bibfnamefont{J.}~\bibnamefont{Lapujoulade}},
  \bibinfo{journal}{Surf. Sci. Rep.} \textbf{\bibinfo{volume}{20}},
  \bibinfo{pages}{191} (\bibinfo{year}{1994}).

\bibitem[{\citenamefont{Saito and
  M{\"u}ller-Krumbhaar}(1981)}]{saito81:_two_coulom}
\bibinfo{author}{\bibfnamefont{Y.}~\bibnamefont{Saito}} \bibnamefont{and}
  \bibinfo{author}{\bibfnamefont{H.}~\bibnamefont{M{\"u}ller-Krumbhaar}},
  \bibinfo{journal}{Phys. Rev. B} \textbf{\bibinfo{volume}{23}},
  \bibinfo{pages}{308} (\bibinfo{year}{1981}).

\end{thebibliography}

\end{document}